\documentclass[preprint2]{aastex}
\usepackage{spr-astr-addons}
\usepackage{graphicx}
\begin{document}

\title{Linkage between Accretion Disks and Blazars}

\shorttitle{Accretion Disks and Blazars}   % if too long for running head
\shortauthors{Bicknell}

\author{Geoffrey V. Bicknell and Jianke Li}
\affil{Research School of Astronomy \& Astrophysics, Australian National University, Cotter Rd., Weston ACT, Australia 2611}

\begin{abstract}
The magnetic field in an accretion disk is estimated assuming that all of the 
angular momentum within prescribed accretion disk radii is removed by a jet. The 
magnetic field estimated at the base of the jet is extrapolated to the blazar 
emission region using a model for a relativistic axisymmetric jet combined with 
some simplifying assumptions based on the relativistic nature of the flow. The 
extrapolated magnetic field is compared with estimates based upon the 
synchrotron and inverse Compton emission from three blazars, MKN~501, MKN~421 
and PKS~2155-304. The magnetic fields evaluated from pure synchrotron self-
Compton models are inconsistent with the magnetic fields extrapolated in this 
way. However, in two cases inverse Compton models in which a substantial part of 
the soft photon field is generated locally agree well, mainly because these 
models imply magnetic field strengths which are closer to being consistent with 
Poynting flux dominated jets. This comparison is based on estimating the mass 
accretion rate from the jet energy flux. Further comparisons along these lines 
will be facilitated by independent estimates of the mass accretion rate in 
blazars and by more detailed models for jet propagation near the black hole.
\end{abstract}



\section{Introduction}
\label{intro}
We believe that the accretion onto black holes and other compact objects is driven by magnetic fields and we can also estimate the required magnetic field (see \S~\ref{B_estimate}). In blazars, we have independent estimates of magnetic fields reasonably close to the black hole at say, $\sim 100-1,000$ gravitational radii. The aim of this paper is to explore the notion that we can relate the magnetic field estimates in these two regions by suitably extrapolating disk magnetic fields to the blazar emission zone. We then examine the conditions under which the extrapolated and estimated fields agree. Some of the physical issues which we examine include those related to Poynting flux dominated jets, the accuracy of specific one-zone inverse Compton models for blazar emission and the accuracy of estimates of the magnetic field strengths in the accretion disk near the black hole.

\section{Magnetic fields and jet power}

\subsection{Disk magnetic field}
\label{s:b_est}

The very existence of a jet or wind in a black hole -- accretion disk system points to the existence of a magnetic stress additional to the conventional $r\phi$ stress identified by \citet{shakura73a}. 
(We adopt $r$, $\phi$ and $z$ as conventional cylindrical polar coordinates with $z=0$ being the disk mid-plane.) \citet{kuncic04a} showed that the channelling of accretion power into either coronal heating or an outflow depends upon the existence of a $\phi z$ stress $\langle t_{\phi z}\rangle ^+$ over the surface of the disk. In principle, this stress can be important even when it is numerically small compared to the $r \phi$ stress $\langle t_{r \phi} \rangle$: $\langle t_{\phi z}\rangle^+$ acts over the surface are a of the disk, whereas $\langle t_{r \phi} \rangle$ acts over the disk height $h$ so that if $\langle t_{\phi z}\rangle^+ \ga (h/r) t_{r\phi}$ then the dynamical effect of $\langle t_{\phi z}\rangle^+$ dominates. 

Let $\dot M_a$ be the disk mass accretion rate and $\tilde v_\phi \approx v_K$ be the approximately Keplerian mean azimuthal velocity\footnote{We utilize the mass weighted averaging of the MHD equations in which direct means are denoted by a bar and mass-weighted means by a tilde. Turbulent components are denoted by a prime and ensemble averages by angular brackets. See \citet{kuncic04a} for details.}, then the full equation for angular momentum transport is:
\begin{equation}
\frac {d}{dr} \left[
\dot M_a \tilde v_\phi r + 
2 \pi r^2 \int_{-h}^{+h} \langle t_{r \phi} \rangle \> dz
\right]  
= \tilde v_\phi r \frac {d \dot M_a}{dr} - 4 \pi r^2 \langle t_{\phi z} \rangle^+
\end{equation}
In principle the mass accretion rate can vary with radius $r$ if there is substantial mass flux deposited into the wind. However, if the wind and/or jet from the accretion disk is Poynting flux dominated then $d \dot M_a /dr \approx 0$. Suppose that we further assume that the $\langle t_{\phi z}\rangle^+$ stress dominates and that the magnetic stresses dominate the kinetic Reynolds stresses ($-\langle \rho v_i^\prime v_j^\prime \rangle$). We then obtain the following estimate for the magnetic term $\langle B_\phi^\prime B_z^\prime \rangle$, \emph{viz.}
\begin{eqnarray}
r^2 \langle B_\phi^\prime B_z^\prime \rangle &=& 
- \frac {d}{dr} \left( \dot M_a r v_K\right) \\ 
\Rightarrow \langle B_\phi^\prime B_z^\prime \rangle &=& 
- \frac {\dot M_a c^5}{2 G^2 M^2} \> \left( \frac {r}{r_g} \right)^{-5/2} \\
\label{e:bphibz1}
&=&
- \frac {2 \pi m_pc^4}{G M \sigma_T} \, \dot m \>
\left( \frac {r}{r_g} \right)^{-5/2}
\label{e:bphibz2}
\end{eqnarray}
where $\dot m$ is the mass accretion rate in units of the Eddington value $\dot M_{\rm Edd} = 4 \pi G M m_p/ c\sigma_T$, $M$ is the mass of the black hole and $r_g = GM/c^2$ is the gravitational radius. Hence we have the following numerical estimate for the strength of the magnetic field:
\begin{equation}
\langle - B_\phi^\prime B_z^\prime \rangle^{1/2} = 3.1 \times 10^4 \, 
M_8^{-1/2} \, \dot m^{1/2} \,
\left( \frac {r}{r_g} \right)^{-5/4}
\end{equation}
where the mass of the black hole is $10^8 M_8 \, M_\odot$. Coincidentally, this is the same dependence of magnetic flux density assumed by \citet{blandford82a} in their paper on centrifugally driven winds from accretion disks. 

\subsection{Jet power}
Let us assume that the jet is Poynting flux dominated and that the base of the jet outflow is confined to radii $r_{\rm in } < r < r_{\rm out}$. We can determine the jet energy flux $F_E$ by integrating the mean Poynting flux
$\langle S_z \rangle \approx - \frac {1}{4 \pi} \langle B_\phi^\prime B_z^\prime \rangle
\tilde v_\phi$ over the cross-sectional area of the jet utilising equation~(\ref{e:bphibz1}) for $\langle B_\phi^\prime B_z^\prime \rangle$. Hence, 
\begin{equation}
F_E = 2 \pi \int \langle S_z \rangle \> r dr \approx 
\frac {\dot M_a c^2}{4} 
\left[ \left( \frac {r_{\rm in }}{r_g} \right)^{-1} - 
\left( \frac {r_{\rm out }}{r_g} \right)^{-1} \right]
\end{equation}
This equation can thus be used to estimate the mass accretion rate given the jet energy flux. Then eliminating $\dot M_a$ from equation~(\ref{e:bphibz1}) we 
have, for the magnetic field as a function of cylindrical radius $r_0$  at the base of the jet:
\begin{equation}
\langle - B_\phi^\prime B_z^\prime \rangle^{1/2} =
\left( \frac {2 F_E}{c} \right)^{1/2} \>
\frac {1}{r_g} \,
\left[ \frac {r_g}{r_{\rm in}} -  \frac {r_g}{r_{\rm out}}
\right]^{-1/2} \,
\left( \frac {r_0}{r_g} \right)^{-5/4}
\end{equation}

The determination of the base magnetic field in terms of the jet can be used in conjunction with a blazar emission model (which gives the jet energy flux) to determine the magnetic field in the jet extrapolated from the accretion disk. This is a valid procedure since the jet energy flux is conserved irrespective of entrainment in the region between the disk and the blazar emission region \citet{bicknell94a}. This is important since the particles in the initially Poynting dominated jet may be entrained en-route. 

\section{Relativistic magnetised jet flow}

In order to connect the magnetic field at the base of the jet to the field further out in the blazar emission region we need to consider the evolution of the field along the Poynting flux dominated jet. In order to do this we consider the equations of relativistic axisymmetric MHD (see \citet{mestel99a}) with a newtonian gravitational field. Let $\Omega$ be the angular velocity of the jet, $\rho$ its rest mass density, ${\bf v}_p=(v_r,0,v_z)$ the poloidal velocity, ${\bf B}_p=(B_r,0,B_z)$ the polodial magnetic field and $\Gamma$ the Lorentz factor. Then we have the following stream line constants:
\begin{eqnarray}
\eta &=& \frac {\rho v_p}{B_p} \nonumber \\
&=& \hbox{Ratio of poloidal rest mass flux } \nonumber \\
&&  \hbox {to poloidal magnetic flux} \\
-\frac {\beta}{4 \pi} &=& - \frac {r B_\phi}{4 \pi} + \eta \Gamma \Omega r^2 
\nonumber \\
&=& \hbox{Rate of angular momentum transport} \nonumber \\
&&  \hbox{per unit poloidal flux tube} \\
\alpha &=& \Omega - \frac {\eta B_\phi}{\rho r} \nonumber \\
&=& \hbox{Angular velocity of field lines} 
\end{eqnarray} 
One of the important surfaces in such a flow is the Alfven surface where the poloidal gas velocity equals the poloidal Alfven speed ($v_p=v_A$). The Alfven surface is also equivalently characterised by $4 \pi \Gamma \eta^2 / \rho = 1$ and $-\beta/4 \pi = \alpha \eta \Gamma r^2$. Denoting the values of dynamical variables at the Alfven surface by a subscript $A$, the azimuthal magnetic field (the dominant perpendicular component) is given by:
\begin{equation}
r B_\phi = - 4 \pi \alpha \eta \Gamma_A r_A^2 \>
\frac {(1 - \Gamma r^2/\Gamma_A r_A^2)}{(1 - \Gamma \rho_A/\Gamma_A \rho)} 
\end{equation}
The asymptotic form of this expression is obtained for $r \gg r_A$, $\rho \ll \rho_A$ and $\Gamma \sim \Gamma_A$, giving 
\begin{equation}
r B_\phi \approx - 4 \pi \alpha \eta \Gamma_A \rho r^2 /\rho_A
\end{equation}
The value of $r B_\phi$ at the disk surface is obtained assuming that $r \ll r_A$ and $\rho \gg \rho_A$ giving 
\begin{equation}
r_0 B_{\phi,0} \approx - 4 \pi \alpha \eta \Gamma_A r_A^2
\label{e:r0b0}
\end{equation}
Hence,
\begin{equation}
\frac {r B_\phi}{r_0 B_\phi} \approx \frac {\rho r^2}{\rho_A r_A^2}
\label{e:rbphi0}
\end{equation}
The other streamline parameter that we require is the ratio of azimuthal to vertical field at the disk surface. The first step is to evaluate the streamline constant 
\begin{equation}
\alpha = \Omega - \frac {\eta B_\phi}{\rho r} = \Omega - 
\left( \frac {v_p}{r}\right) \, \left( \frac {B_\phi}{B_p} \right)
\end{equation}
Assuming that $B_\phi$ is at most of order $B_p$ and that the initial poloidal velocity is much less than the Keplerian velocity ($v_p \ll v_{\rm K,0} = r \Omega$) we have
\begin{equation}
\alpha \approx \Omega
\end{equation}
Then, dividing equation~(\ref{e:r0b0}) by $r_0 B_{z,0}$ and using $B_{z,0} = \eta^{-1} \rho_0 v_{z,0}$ gives
\begin{equation}
\frac {B_{\phi,0}}{B_{z,0}} \approx - \left( \frac {v_{K,0}}{v_A} \right) \,
\left( \frac {\rho_A v_A r_A^2}{\rho_0 v_{z,0} r_0^2} \right)
\label{e:bphibz}
\end{equation}
In both equations~(\ref{e:r0b0}) and (\ref{e:rbphi0}) we have the similar terms $\rho r^2 /\rho_A r_A^2$ and $\rho_A v_A r_A^2/ \rho_0 v_{z,0} r_0^2$ which we can treat approximately as follows. Let the equation of a poloidal stream line by $r = R(z)$. The continuity equation tells us that
\begin{equation}
\frac {d}{dz} \int_0^{R(z)} \rho(r,z) v_z(r,z) r \> dr =0
\end{equation} 
If we assume that over a jet cross-section the density and vertical velocity are constant, \emph{i.e.} $\rho(r,z) \approx \rho(z)$ and $v_z(r,z) \approx v_z(z)$, then
\begin{equation}
\rho(z) v_z(z) R^2(z) \approx \hbox{constant}
\end{equation}
Thus along any poloidal streamline, $\rho v_z r^2 \approx \hbox{constant}$. Consequently
\begin{eqnarray}
\frac {\rho r^2}{\rho_A r_A^2} &\approx& \frac {v_A}{v_z} \\
\frac {\rho_A v_A r_A^2}{\rho_0 v_{z,0} r_0^2} &\approx& 1
\end{eqnarray}
Furthermore, we assume that by the time the flow has reached the Alfven surface  it is already relativistic although possibly not with a high Lorentz factor. Hence, we assume that $v_A \approx c$. Also, in the asymptotic regime where $\Gamma \gg 1$, $v_z \approx c$ to an even better approximation.
Therefore, in the asymptotic region 
\begin{equation}
\frac {r B_\phi}{r_0 B_\phi} \approx \frac {v_A}{v_z} \approx 1
\label{e:rbphi_final}
\end{equation}
The ratio of the azimuthal and vertical components of magnetic field in the disk is:
\begin{equation}
\frac {B_{\phi,0}}{B_{z,0}} \approx - \frac {v_{K,0}}{v_A} 
\approx - \frac {v_{K,0}}{c}
\end{equation}
Given this estimate of the ratio of toroidal to vertical field components at the base of the jet we can use the estimate for 
$\langle B_{\phi,0}B_{z,0}\rangle^+$ on the accretion disk surface to determine the asymptotic azimuthal field using equation~(\ref{e:rbphi_final}). 
The result is:
\begin{equation}
r B_\phi \approx \left( \frac {2 F_E}{c} \right)^{1/2} \,
\left[ \left(\frac {r_g}{r_{\rm in }}\right) - 
\left( \frac {r_g}{r_{\rm out}} \right)\right]^{-1/2} \,
\left( \frac {r_0}{r_g} \right)^{-1/2}
\end{equation}
where $r_0$ refers to the value of $r$ at the base of the streamline. 
In order to compare this estimate of magnetic field with the one-zone models which invoke a single magnetic field strength we define an area weighted flux density
\begin{equation}
\bar B_\phi = \frac {2 \pi \int_0^{R_{\rm jet}} r B_\phi \> dr}{\pi R_{\rm jet}^2}
\end{equation}
where $R_{\rm jet}$ is the radius of the jet in the blazar emission region. In order to carry out this integral one needs to integrate$(r_0/r_g)^{-1/2}$ with respect to the streamline radius $r$ across the jet in the blazar region; this in turn requires that we know the relationship between the streamline radius and the radius at the base of the streamline, i.e. $r(r_0)$. We make the simplifying assumption that the expansion is linear. This should not be a great source of error since the integrand only depends weakly on $r_0$. With this assumption and with $x = r_{\rm in}/r_{\rm out}$ as the ratio of inner to outer jet radii:
\begin{eqnarray}
\bar B_\phi &\approx& 2^{5/2} \left(\frac{F_E}{c} \right)^{1/2} \, f(x) \,
 \frac {1}{R_{\rm jet}} \label{e:b_lab} \\
\hbox{where} \qquad f(x) &=& \left[ \frac {x(1-x^{1/2})}{1+x^{1/2}}\right]^{1/2}
\label{e:bphi_lab}
\end{eqnarray}
The function $f(x)$ varies by a factor of 2 from its mean value $\approx 0.24$ for $0.019 \la x \la 0.94$ so that the precise value of $x$ is unimportant. However, for illustration we consider two different values in the following.

The magnetic field estimate in equation~(\ref{e:b_lab}) represents the magnetic field in the laboratory frame. For comparison with blazar estimates the magnetic field in the plasma rest frame is obtained by dividing this estimate by the Lorentz factor $\Gamma$.    

\section{Comparison with models of blazar emission}

Our aim is to compare the estimate of the magnetic field using the above theory with the values derived from inverse Compton models of blazar high energy emission. Inverse Compton emission results from the scattering of ``soft'' photons by high energy electrons and one may distinguish three qualitatively different models for blazar emission:
\begin{enumerate}
\item \emph{Exernal inverse Compton (EIC)} models in which the soft photons originate from outside the region of high energy electrons.
\item \emph{Synchroton Self Compton (SSC)} models in which the soft photons are produced by the same population of high energy electrons which scatter them.
\item \emph{Local Inverse Compton (LIC)} models in which the soft photons originate from a region which is adjacent to the region of high energy electrons. However, the regions may be physically related. For example, one can envisage a region of jet containing embedded shocks. The high energy electrons would then be in a sub-volume of a larger volume which is the source of soft photons. 
\end{enumerate}

In the literature the distnction between LIC emission and EIC emission is not usually made. However, we think that the distinction is useful in view of the close spatial and physical relationship that may exist between the source of soft photons and the scattering region. Examples of LIC emission models include the decelerating flow models proposed by \citet{georganopoulos06a} and the `blob in jet' model of \citet{katarzynski01a}.

\begin{table*}[h!]
\begin{center}
\begin{tabular}{| c | c | c | c | c | c | c | c | c | c |}
\hline
\bf Source & \bf Blazar & $R_{\rm jet}$ & $\Gamma_{\rm jet}$ & $K$ &
$B_{\rm IC}$ & $F_E$  & $r_{\rm in }/ r_{\rm out}$ &$B_{\rm ext}$ & 
 $B_{\rm ext}/B_{\rm IC}$ \\
       & \bf Model      & $10^{16} \> \rm cm$ &      & $\rm cm^{-3}$  & G             & 
$10^{43} \> \rm ergs \> s^{-1}$&   &  G            & \\
\hline
MKN 501 & SSC$^1$ & 0.53 & 10 & $4.3 \times 10^3$ & 0.06 & 7.0  & 0.1 & 1.3 & 22 \\
MKN 501 & SSC$^1$ & 0.53 & 10 & $4.3 \times 10^3$ & 0.06 & 7.0  & 0.5 & 1.7 & 28 \\
MKN 501 & LIC$^2$ & 1.0 & 7  & 100 & 0.15 & 0.33 & 0.1 & 0.19 & 1.3 \\
MKN 501 & LIC$^2$ & 1.0 & 7  & 100 & 0.15 & 0.33 & 0.5 & 0.25 & 1.7 \\
\hline
MKN 421 & SSC$^3$ & 1.0 & 10 & $4 \times 10^4$ & 0.06 & 100 & 0.1 & 2.4 & 39 \\
MKN 421 & SSC$^3$ & 1.0 & 10 & $4 \times 10^4$ & 0.06 & 100 & 0.5 & 3.0 & 50 \\
\hline
PKS 2155-304 & LIC$^4$ & 0.65 & 25 & 160 & 0.12 & 0.68 & 0.1 & 0.12 & 0.8 \\
PKS 2155-304 & LIC$^4$ & 0.65 & 25 & 160 & 0.15 & 0.68 & 0.5 & 0.15 & 1.0 \\
\hline
\end{tabular}
\end{center}
\caption{Comparison of the extrapolated magnetic field $B_{\rm ext}$ and the inverse-Compton estimated magnetic field $B_{\rm IC}$ for three blazars. The jet radius, Lorentz factor, number density and $B_{\rm IC}$ are provided by the cited models; the jet energy flux is inferred from those models. The extrapolated magnetic field is calculated using the above theory and the ratio of extrapolated to inverse Compton magnetic field is given in the last column of the table. References: 1: \citet{bicknell01a}; 2: \citet{katarzynski01a}; 3:\citet{maraschi99a}; 4: \citet{aharonian05b} Model 1 }
\label{t:blazars}
\end{table*}

Table~\ref{t:blazars} summarises the comparison of blazar and extrapolated accretion disk estimates for three blazars. Different groups have independently estimated blazar parameters using different models (SSC and LIC) for MKN~501. In the comparisons we have used two different values of $x=r_{\rm in}/r_{\rm out}$; however, as indicated earlier this does not make great deal of difference to the extrapolated magnetic field estimates. We have also indicated which of the models are SSC and which are LIC; the last column of the table indicates the ratio of the extrapolated magnetic field  to the magnetic field estimated from the particular blazar model. 

\section{Discussion}
One of the obvious conclusions from Table~\ref{t:blazars} is that the SSC models do not fare very well in estimating the magnetic field in the blazar region. On the other hand the two LIC estimates of magnetic flux density agree well with the extrapolated magnetic field. Whilst such an agreement is welcome it is not especially surprising. The estimate of the magnetic field in equation~(\ref{e:bphi_lab}) is based on a Poynting flux dominated jet which remains so in the asymptotic regime -- consistent with the original non-relativistic solution of \citet{blandford82a}. It is well known that pure SSC models yield low magnetic energy densities and high particle energy densities that are incompatible with an important Poynting flux. The LIC models on the other hand effectively reduce the particle energy density and increase the magnetic energy density as a consequence of the larger flux of soft photons. The LIC models therefore imply comparable values of magnetic flux density and energy density, which are more compatible with a substantial component of Poynting flux. Hence, the agreement with the extrapolated field is not surprising. In order to provide a more stringent test of the extrapolation technique developed here it is desirable to have an independent estimate of the accretion rate (rather than one based on the energy flux) which can be used to estimate the base magnetic field (using equation~(\ref{e:bphibz2})) independently of the energy flux. This could be provided, for example, by the X-ray and/or optical luminosity associated with the accretion disk.

On the other hand, the reasonable comparison between extrapolated and estimated magnetic fields bodes well for a continuation of this approach in which several of the simplifying assumptions can be relaxed. In particular the gravitational field of the black hole can be incorporated into a fully general relativistic treatment, a small mass flux can be incorporated and the consistent development of the model for the flow near the black hole may also be included.

\begin{acknowledgements}
We would like to acknowledge invaluable collaborations with Zdenka Kuncic and Stefan Wagner.
\end{acknowledgements}

% BibTeX users please use the following style file
%\bibliographystyle{spr-mp-nameyear}
%\bibliography{gvbrefs}

\begin{thebibliography}{}
\ifx \bisbn   \undefined \def \bisbn  #1{ISBN #1}   \fi
\ifx \binits  \undefined \def \binits#1{#1} \fi
\ifx \bauthor  \undefined \def \bauthor#1{#1} \fi
\ifx \batitle  \undefined \def \batitle#1{#1} \fi
\ifx \bjtitle  \undefined \def \bjtitle#1{#1} \fi
\ifx \bvolume  \undefined \def \bvolume#1{#1} \fi
\ifx \byear  \undefined \def \byear#1{#1} \fi
\ifx \bissue  \undefined \def \bissue#1{#1} \fi
\ifx \bfpage  \undefined \def \bfpage#1{#1} \fi
\ifx \blpage  \undefined \def \blpage #1{#1} \fi
\ifx \burl  \undefined \def \burl#1{#1} \fi
\ifx \binterref  \undefined \def \binterref#1{#1} \fi
\ifx \betal  \undefined \def \betal#1{#1} \fi
\ifx \binstitute  \undefined \def \binstitute#1{#1} \fi
\ifx \bctitle  \undefined \def \bctitle#1{#1} \fi
\ifx \beditor  \undefined \def \beditor#1{#1} \fi
\ifx \bpublisher  \undefined \def \bpublisher#1{#1} \fi
\ifx \bbtitle  \undefined \def \bbtitle#1{#1} \fi
\ifx \bedition  \undefined \def \bedition#1{#1} \fi
\ifx \bseriesno  \undefined \def \bseriesno#1{#1} \fi
\ifx \blocation  \undefined \def \blocation#1{#1} \fi
\ifx \bsertitle  \undefined \def \bsertitle#1{#1} \fi
\ifx \bsnm \undefined \def \bsnm#1{#1} \fi
\ifx \bsuffix \undefined \def \bsuffix#1{#1} \fi
\ifx \bparticle \undefined \def \bparticle#1{#1} \fi
\ifx \barticle \undefined \def \barticle#1{#1} \fi
\ifx \botherref \undefined \def \botherref #1{#1} \fi
\ifx \url \undefined \def \url#1{#1} \fi
\ifx \bchapter \undefined \def \bchapter#1{#1} \fi
\ifx \bbook \undefined \def \bbook#1{#1} \fi
\ifx \bcomment \undefined \def \bcomment#1{#1} \fi
\ifx \protect\citeauthoryear \undefined \def \protect\citeauthoryear#1{#1} \fi
\ifx \oauthor \undefined \def \oauthor#1{#1} \fi
\def \endbibitem {}

\bibitem[\protect\citeauthoryear{{Shakura} and {Sunyaev}}{1973}]{shakura73a}
\begin{barticle}
\bauthor{\bsnm{{Shakura}},~\binits{N.I.}},
  \bauthor{\bsnm{{Sunyaev}},~\binits{R.A.}}:
\batitle{{Black holes in binary systems. Observational appearance.}}
\bjtitle{A\&A} \bvolume{24},  \bfpage{337}--\blpage{355} (\byear{1973})
\end{barticle}
\endbibitem

\bibitem[\protect\citeauthoryear{{Kuncic} and {Bicknell}}{2004}]{kuncic04a}
\begin{barticle}
\bauthor{\bsnm{{Kuncic}},~\binits{Z.}},
  \bauthor{\bsnm{{Bicknell}},~\binits{G.V.}}:
\batitle{Dynamics and Energetics of Turbulent, Magnetized Disk Accretion around
  Black Holes: a First-Principles Approach to Disk--Corona--Outflow Coupling}.
\bjtitle{ApJ} \bvolume{616},  \bfpage{669}--\blpage{687} (\byear{2004})
\end{barticle}
\endbibitem

\bibitem[\protect\citeauthoryear{Blandford and Payne}{1982}]{blandford82a}
\begin{barticle}
\bauthor{\bsnm{Blandford},~\binits{R.D.}},
  \bauthor{\bsnm{Payne},~\binits{D.G.}}
\bjtitle{MNRAS} \bvolume{199},  883 (\byear{1982})
\end{barticle}
\endbibitem

\bibitem[\protect\citeauthoryear{Bicknell}{1994}]{bicknell94a}
\begin{barticle}
\bauthor{\bsnm{Bicknell},~\binits{G.V.}}:
\batitle{On the Relationship between BL Lacertae Objects and Fanaroff-Riley I
  Radio Galaxies}.
\bjtitle{ApJ} \bvolume{422},  542 (\byear{1994})
\end{barticle}
\endbibitem

\bibitem[\protect\citeauthoryear{Mestel}{1999}]{mestel99a}
\begin{bbook}
\bauthor{\bsnm{Mestel},~\binits{L.}}:
\bbtitle{Stellar Magnetism}. \bsertitle{The International Series of Monographs
  on Physics}. \bpublisher{Clarendon Press}, \blocation{Oxford} (\byear{1999})
\end{bbook}
\endbibitem

\bibitem[\protect\citeauthoryear{{Kazanas} and
  {Georganopoulos}}{2006}]{georganopoulos06a}
\begin{botherref}
\oauthor{\bsnm{{Kazanas}}~\binits{D}},
  \oauthor{\bsnm{{Georganopoulos}}~\binits{M}}:
{Decelerating Flows in TeV Blazars: A Resolution to the BL Lacertae-FR I
  Unification Problem}.
In: {Miller}, H.R., {Marshall}, K., {Webb}, J.R., {Aller}, M.F. (eds.) Blazar
  Variability Workshop II: Entering the GLAST Era. Astronomical Society of the
  Pacific Conference Series,  vol. 350, p. 124. (2006)
\end{botherref}
\endbibitem

\bibitem[\protect\citeauthoryear{{Katarzy{\'n}ski}, {Sol}, and
  {Kus}}{2001}]{katarzynski01a}
\begin{barticle}
\bauthor{\bsnm{{Katarzy{\'n}ski}},~\binits{K.}},
  \bauthor{\bsnm{{Sol}},~\binits{H.}}, \bauthor{\bsnm{{Kus}},~\binits{A.}}:
\batitle{{The multifrequency emission of Mrk 501. From radio to TeV
  gamma-rays}}.
\bjtitle{A\&A} \bvolume{367},  \bfpage{809}--\blpage{825} (\byear{2001}).
\end{barticle}
\endbibitem

\bibitem[\protect\citeauthoryear{{Bicknell}, {Wagner}, and
  {Groves}}{2001}]{bicknell01a}
\begin{botherref}
\oauthor{\bsnm{{Bicknell}}~\binits{GV}}, \oauthor{\bsnm{{Wagner}}~\binits{SJ}},
  \oauthor{\bsnm{{Groves}}~\binits{B}}:
{Gamma-ray Emission from Active Galactic Nuclei-An Overview}.
In: Aharonian, F.A., V\"olk, H.J. (eds.) International Symposium on High Energy
  Gamma Ray Astronomy.  vol. 558, p. 261. A.I.P. (2001)
\end{botherref}
\endbibitem

\bibitem[\protect\citeauthoryear{{Maraschi}
  \textit{et~al.}}{1999}]{maraschi99a}
\begin{barticle}
\bauthor{\bsnm{{Maraschi}},~\binits{L.}},
  \bauthor{\bsnm{{Fossati}},~\binits{G.}},
  \bauthor{\bsnm{{Tavecchio}},~\binits{F.}},
  \bauthor{\bsnm{{Chiappetti}},~\binits{L.}},
  \bauthor{\bsnm{{Celotti}},~\binits{A.}},
  \bauthor{\bsnm{{Ghisellini}},~\binits{G.}},
  \bauthor{\bsnm{{Grandi}},~\binits{P.}}, \bauthor{\bsnm{{Pian}},~\binits{E.}},
  \bauthor{\bsnm{{Tagliaferri}},~\binits{G.}},
  \bauthor{\bsnm{{Treves}},~\binits{A.}},
  \bauthor{\bsnm{{Breslin}},~\binits{A.C.}},
  \bauthor{\bsnm{{Buckley}},~\binits{J.H.}},
  \bauthor{\bsnm{{Carter-Lewis}},~\binits{D.A.}},
  \bauthor{\bsnm{{Catanese}},~\binits{M.}},
  \bauthor{\bsnm{{Cawley}},~\binits{M.F.}},
  \bauthor{\bsnm{{Fegan}},~\binits{D.J.}},
  \bauthor{\bsnm{{Fegan}},~\binits{S.}},
  \bauthor{\bsnm{{Finley}},~\binits{J.}},
  \bauthor{\bsnm{{Gaidos}},~\binits{J.}}, \bauthor{\bsnm{{Hall}},~\binits{T.}},
  \bauthor{\bsnm{{Hillas}},~\binits{A.M.}},
  \bauthor{\bsnm{{Krennrich}},~\binits{F.}},
  \bauthor{\bsnm{{Lessard}},~\binits{R.W.}},
  \bauthor{\bsnm{{Masterson}},~\binits{C.}},
  \bauthor{\bsnm{{Moriarty}},~\binits{P.}},
  \bauthor{\bsnm{{Quinn}},~\binits{J.}}, \bauthor{\bsnm{{Rose}},~\binits{J.}},
  \bauthor{\bsnm{{Samuelson}},~\binits{F.}},
  \bauthor{\bsnm{{Weekes}},~\binits{T.C.}},
  \bauthor{\bsnm{{Urry}},~\binits{C.M.}},
  \bauthor{\bsnm{{Takahashi}},~\binits{T.}}:
\batitle{Simultaneous X-Ray and TEV Observations of a Rapid Flare from
  Markarian 421}.
\bjtitle{ApJL} \bvolume{526}, ~81 (\byear{1999})
\end{barticle}
\endbibitem

\bibitem[\protect\citeauthoryear{{Aharonian}
  \textit{et~al.}}{2005}]{aharonian05b}
\begin{barticle}
\bauthor{\bsnm{{Aharonian}},~\binits{F.}},
  \bauthor{\bsnm{{Akhperjanian}},~\binits{A.G.}},
  \bauthor{\bsnm{{Aye}},~\binits{K.M.}},
  \bauthor{\bsnm{{Bazer-Bachi}},~\binits{A.R.}},
  \bauthor{\bsnm{{Beilicke}},~\binits{M.}},
  \bauthor{\bsnm{{Benbow}},~\binits{W.}},
  \bauthor{\bsnm{{Berge}},~\binits{D.}},
  \bauthor{\bsnm{{Berghaus}},~\binits{P.}},
  \bauthor{\bsnm{{Bernl{\"o}hr}},~\binits{K.}},
  \bauthor{\bsnm{{Bolz}},~\binits{O.}},
  \bauthor{\bsnm{{Boisson}},~\binits{C.}},
  \bauthor{\bsnm{{Borgmeier}},~\binits{C.}},
  \bauthor{\bsnm{{Breitling}},~\binits{F.}},
  \bauthor{\bsnm{{Brown}},~\binits{A.M.}}, \bauthor{\bsnm{{Bussons
  Gordo}},~\binits{J.}}, \bauthor{\bsnm{{Chadwick}},~\binits{P.M.}},
  \bauthor{\bsnm{{Chitnis}},~\binits{V.R.}},
  \bauthor{\bsnm{{Chounet}},~\binits{L.M.}},
  \bauthor{\bsnm{{Cornils}},~\binits{R.}},
  \bauthor{\bsnm{{Costamante}},~\binits{L.}},
  \bauthor{\bsnm{{Degrange}},~\binits{B.}},
  \bauthor{\bsnm{{Djannati-Ata{\"i}}},~\binits{A.}},
  \bauthor{\bsnm{{Drury}},~\binits{L.O.}},
  \bauthor{\bsnm{{Ergin}},~\binits{T.}},
  \bauthor{\bsnm{{Espigat}},~\binits{P.}},
  \bauthor{\bsnm{{Feinstein}},~\binits{F.}},
  \bauthor{\bsnm{{Fleury}},~\binits{P.}},
  \bauthor{\bsnm{{Fontaine}},~\binits{G.}},
  \bauthor{\bsnm{{Funk}},~\binits{S.}},
  \bauthor{\bsnm{{Gallant}},~\binits{Y.A.}},
  \bauthor{\bsnm{{Giebels}},~\binits{B.}},
  \bauthor{\bsnm{{Gillessen}},~\binits{S.}},
  \bauthor{\bsnm{{Goret}},~\binits{P.}}, \bauthor{\bsnm{{Guy}},~\binits{J.}},
  \bauthor{\bsnm{{Hadjichristidis}},~\binits{C.}},
  \bauthor{\bsnm{{Hauser}},~\binits{M.}},
  \bauthor{\bsnm{{Heinzelmann}},~\binits{G.}},
  \bauthor{\bsnm{{Henri}},~\binits{G.}},
  \bauthor{\bsnm{{Hermann}},~\binits{G.}},
  \bauthor{\bsnm{{Hinton}},~\binits{J.A.}},
  \bauthor{\bsnm{{Hofmann}},~\binits{W.}},
  \bauthor{\bsnm{{Holleran}},~\binits{M.}},
  \bauthor{\bsnm{{Horns}},~\binits{D.}}, \bauthor{\bsnm{{de
  Jager}},~\binits{O.C.}}, \bauthor{\bsnm{{Jung I.}},~},
  \bauthor{\bsnm{{Kh{\'e}lifi}},~\binits{B.}},
  \bauthor{\bsnm{{Komin}},~\binits{N.}},
  \bauthor{\bsnm{{Konopelko}},~\binits{A.}},
  \bauthor{\bsnm{{Latham}},~\binits{I.J.}}, \bauthor{\bsnm{{Le
  Gallou}},~\binits{R.}}, \bauthor{\bsnm{{Lemoine}},~\binits{M.}},
  \bauthor{\bsnm{{Lemi{\`e}re}},~\binits{A.}},
  \bauthor{\bsnm{{Leroy}},~\binits{N.}}, \bauthor{\bsnm{{Lohse}},~\binits{T.}},
  \bauthor{\bsnm{{Marcowith}},~\binits{A.}},
  \bauthor{\bsnm{{Masterson}},~\binits{C.}},
  \bauthor{\bsnm{{McComb}},~\binits{T.J.L.}}, \bauthor{\bsnm{{de
  Naurois}},~\binits{M.}}, \bauthor{\bsnm{{Nolan}},~\binits{S.J.}},
  \bauthor{\bsnm{{Noutsos}},~\binits{A.}},
  \bauthor{\bsnm{{Orford}},~\binits{K.J.}},
  \bauthor{\bsnm{{Osborne}},~\binits{J.L.}},
  \bauthor{\bsnm{{Ouchrif}},~\binits{M.}},
  \bauthor{\bsnm{{Panter}},~\binits{M.}},
  \bauthor{\bsnm{{Pelletier}},~\binits{G.}},
  \bauthor{\bsnm{{Pita}},~\binits{S.}}, \bauthor{\bsnm{{Pohl}},~\binits{M.}},
  \bauthor{\bsnm{{P{\"u}hlhofer}},~\binits{G.}},
  \bauthor{\bsnm{{Punch}},~\binits{M.}},
  \bauthor{\bsnm{{Raubenheimer}},~\binits{B.C.}},
  \bauthor{\bsnm{{Raue}},~\binits{M.}}, \bauthor{\bsnm{{Raux}},~\binits{J.}},
  \bauthor{\bsnm{{Rayner}},~\binits{S.M.}},
  \bauthor{\bsnm{{Redondo}},~\binits{I.}},
  \bauthor{\bsnm{{Reimer}},~\binits{A.}},
  \bauthor{\bsnm{{Reimer}},~\binits{O.}},
  \bauthor{\bsnm{{Ripken}},~\binits{J.}},
  \bauthor{\bsnm{{Rivoal}},~\binits{M.}}, \bauthor{\bsnm{{Rob}},~\binits{L.}},
  \bauthor{\bsnm{{Rolland}},~\binits{L.}},
  \bauthor{\bsnm{{Rowell}},~\binits{G.}},
  \bauthor{\bsnm{{Sahakian}},~\binits{V.}},
  \bauthor{\bsnm{{Saug{\'e}}},~\binits{L.}},
  \bauthor{\bsnm{{Schlenker}},~\binits{S.}},
  \bauthor{\bsnm{{Schlickeiser}},~\binits{R.}},
  \bauthor{\bsnm{{Schuster}},~\binits{C.}},
  \bauthor{\bsnm{{Schwanke}},~\binits{U.}},
  \bauthor{\bsnm{{Siewert}},~\binits{M.}}, \bauthor{\bsnm{{Sol}},~\binits{H.}},
  \bauthor{\bsnm{{Steenkamp}},~\binits{R.}},
  \bauthor{\bsnm{{Stegmann}},~\binits{C.}},
  \bauthor{\bsnm{{Tavernet}},~\binits{J.P.}},
  \bauthor{\bsnm{{Th{\'e}oret}},~\binits{C.G.}},
  \bauthor{\bsnm{{Tluczykont}},~\binits{M.}}, \bauthor{\bsnm{{van der
  Walt}},~\binits{D.J.}}, \bauthor{\bsnm{{Vasileiadis}},~\binits{G.}},
  \bauthor{\bsnm{{Vincent}},~\binits{P.}},
  \bauthor{\bsnm{{Visser}},~\binits{B.}},
  \bauthor{\bsnm{{V{\"o}lk}},~\binits{H.J.}},
  \bauthor{\bsnm{{Wagner}},~\binits{S.J.}}:
\batitle{{H.E.S.S. observations of PKS 2155-304}}.
\bjtitle{\aap} \bvolume{430},  \bfpage{865}--\blpage{875} (\byear{2005}).
\end{barticle}
\endbibitem

\end{thebibliography}

%\begin{thebibliography}{}
%\bibitem[\protect\citeauthoryear{Author(s)}{year}]{label}
% etc

\end{document}